# Temporal evolution of mechanical properties of skeletal tissue regeneration in rabbits. An experimental study.

Didier Moukoko MD[1-2], Martine Pithioux PhD[1], Patrick Chabrand PU[1].

[1]Laboratoire d'Aérodynamique et de Biomécanique du Mouvement (LABM)
163, Avenue de Luminy, Case Postale 918
13288 MARSEILLE Cedex 09, France

[2]Laboratoire de Chirurgie Expérimentale, Faculté de Médecine, Université Montpellier I
4 Boulevard Henri IV
34000 Montpellier, France

**Abbreviated title**: Mechanical properties of skeletal tissue regeneration

**Corresponding author**:

Martine Pithioux

Laboratoire d'Aérodynamique et de Biomécanique du Mouvement, Université de la méditerranée, 163 avenue de Luminy, case 918

13288 Marseille, France.

Tel. 334 912 661 77 Fax. 334 914 116 91

E-mail address: martine.pithioux@univmed.fr

**Abstract**

Various mathematical models represent the effects of local mechanical environment on the regulation of skeletal regeneration. Their relevance relies on an accurate description of the evolving mechanical properties of the regenerating tissue. The object of this study was to develop an experimental model which made it possible to characterize the temporal evolution of the structural and mechanical properties during unloaded enchondral osteogenesis in the New Zealand rabbit, a standard animal model for studies of osteogenesis and chondrogenesis. A 25mm segment of tibial diaphysis was removed sub-periosteally from rabbits. The defect was repaired by the preserved periosteum. An external fixator was applied to prevent mechanical loading during osteogenesis. The regenerated skeletal tissues were studied by CT scan, histology and mechanical tests. The traction tests between 7 to 21 days post-surgery were done on formaldehyde-fixated tissue allowing to obtain force/displacement curves. The viscoelastic properties of the regenerating skeletal tissues were visualized throughout the repair process.

**Key Terms**: Biomechanics, Bone regeneration, Osteogenesis, Periosteum, Physiology

## 1) Introduction

In the literature, skeletal tissue regeneration has been extensively studied. Its general process involves mesenchymal precursor cells ([1]). which first proliferate to fill the tissue volume ([2]), then differentiate into mature tissue endowed with remodelling properties ([3;4]). These cascades of cellular and molecular events follow programmed time sequences modulated by environmental factors. Among them, biological signals and the local mechanical environment interact with each step of the cascade of cell events involved in tissue regeneration. These interactions led to the theoretical concepts of mechanobiology ([4;5]). The stress and strains applied to the precursor cells induce specific biological responses at all levels of the regenerating skeletal structure, from the molecular level of its extra-cellular matrix ([6;7]) to its macroscopic morphology ([4;8]).

Various numerical model studies have been used to analyse the temporal evolution of mechanical properties of multipotent mesenchymal tissue ([4;9-11]). Two types of mechanical behaviour have been considered: elastic ([4;5;12;13]); and poroelastic ([4;5;9-13]). These models have been confirmed by experiments which are not realised on a regenerating tissue.

The New Zealand white rabbit is a standard experimental model for study of osteogenesis and chondrogenesis. However, the temporal evolution of the mechanical properties of these regenerating tissues has not been studied. This work aims to characterize the tensile properties of regenerating skeletal tissue at 1, 2 and 3 weeks in a rabbit tibial defect model.

Our experimental model of skeletal tissue regeneration is based on periosteal properties. Periosteal stripping from the bone surface induces a cascade of cellular and molecular events which brings mesenchymal precursor cells to the surgical bed ($^{14}$). They will then proliferate before differentiating.

The *in vitro* test measured the temporal evolution of the mechanical behaviour of this skeletal tissue from the mesenchymal stages until ossification into primary bone. CT scan and histological study completed the traction tests.

## 2) Materials and Methods

Fourteen three-month old New Zealand rabbits weighing 2.5 kg and skeletally immature were used as animal models (INRA-ENSA Montpellier France). The surgery was performed in an accredited experimental surgery laboratory of Montpellier Medical School, in accordance with French regulations on animal care and use of laboratory animals. Two rabbits died during anesthesia; however none of the animals suffered postoperative complication such as pin tract or wound infection. In the postoperative period, the animals ate and walked normally. No animal died postoperatively.

2 - 1)   Animal model

The anesthesia was obtained by premedication with acepromazine 1%, (Labo. CEVA) 10 mg IM, followed by IV continuous infusion of 2% xylazine (Labo. BAYER) 0.27 mg/mn, and 5% ketamine (Labo. MERIAL ) 1.14 mg/mn.

Under strict surgical aseptic conditions the medial third of one tibia was exposed via a medial approach. The periosteal sheet was incised longitudinally on the lateral side of the tibia in order to preserve its vascular connections with the saphenous bundle. The periosteum was elevated from the entire circumference of the bone segment. In order to maintain the leg length and axis, a mono-lateral external fixator (Orthofix M 111) was applied and secured to the bone by 4 half pins (diameter 2.5 mm). A 25 mm long bone segment was cut off and removed from the medial third of the tibia. The periosteal mantel was carefully closed back over the segmental bone loss covering the edge of the bone section in order to initiate the skeletal regeneration and act as a barrier against adjacent soft tissue interposition.

The skin was closed around the pins on the medial side off the hind limb. After surgery the animal was restricted to ambulation in its cage. The external fixator allowed the bypass of the mechanical load through the regenerating tissues during the locomotion of the animals (Figure 1).

After the prescribed healing time, the rabbits were sacrificed by an overdose of pentobarbital at chronological dates: Day 7, 14 and 21. The operated tibiae were explanted in order to analyse the regenerated tissues at these three stages of maturation. The tissue samples included the regenerated zones (25 mm) in continuity with their bone attachments on either side, from the knee joint up to 1 cm beyond the most distal pin fixation (Figure 2). Twelve regenerating tibiae (5 obtained at 7 days, 4 at 14 days, 3 at 21 days) and one healthy bone were available for mechanical analysis.

2 - 2) *In vitro* experiments

**Image acquisition**:

Longitudinal CT scan images of the regenerating tissue samples were obtained (Light speed, General Electric, 0.6 mm between each image). Imaging showed the shape and contours of the regenerating structures and illustrated the evolution of their mineralization. The time required for image acquisition did not exceed five minutes, thus preventing dehydration of the tissues.

**Mechanical evaluation**

Non destructive hysteresis and relaxation tests were performed followed by quasi-static tensile tests, up to failure. Traction tests were chosen to analyze mechanical properties of the samples rather than compression tests to avoid problems of buckling

Thirteen fresh samples (5 obtained at 7 days, 4 at 14 days, 3 at 21 days and 1 healthy bone) were preserved in formaldehyde ([15]). They all were kept in similar preservation and experimental conditions. The healthy bone was tested to evaluate qualitatively its mechanical behaviour. The 2.5 mm pins of the external fixator were pushed through the bone to serve as an anchor to the traction device.

Twined cables (Ø 0.8) were used to tie the fixators (Figure 2). Tensile experiments were carried out using a common tensile device - Instron. The displacement of the lower traverse beam was measured by an LVDT sensor (Linear Variable Differential Transformer), attached to the machine frame. The tensile load was measured by a load cell (Instron) on the upper traverse beam, with measurement error of ±2.5N and maximum load to 1 kN. The displacement rate of the lower traverse beam was 10mm/min. In order to describe the strain of the tissue samples rather than changes of length that would include the progressive tensioning of cables in the traction device (Figure 2), an extensometer (Instron) was used.

Cyclic and relaxation loading regimens were used to characterize viscous and plastic effects. We prescribed three loading cycles, the strain rate loading was 0.4 $min^{-1}$. In relaxation experiments, displacements were prescribed and the force variations were measured during at the very least 30 seconds.

Then, on the same device, quasi-static tensile experiments were performed.

**Statistics:**

Because of the weak number of samples, a non-parametric Mann Whitney test was carried out to compare the groups of samples at 7, 14 and 21 days. The comparisons were realized in pairs between the groups of failure force and between the groups of failure displacement.

**Histology:**

After mechanical testing, a histological study of the samples was performed. Twelve transverse sections were cut every 2 mm. Hematoxylin-Eosin-Saffron (HES) and Giemsa stains were used. Counting cells in a grid allowed an estimate of the proportion of the different cell types in the tissues.

**3) Results:**

3 - 1) Anatomical results

In all animals, the bone defect was successfully repaired by the regenerating tissue, producing structural continuity between opposing bone sections. The overall outline of the regenerating tissue reproduced roughly the morphology of the removed bone segment (Figure 2). CT scan reconstructions analyzed the axial alignment, and the outline of the native bone extremities appeared clearly on X-ray images (Figure 3). With time, the mineralization of the regenerating structure proceeded by an increase of density within its structure, particularly in its periphery where the cortical bone was expected to appear in the future.

3 - 2) Mechanical testing

Figure 4a shows the force/displacement curves obtained from the cyclic tests, with the residual strain at the end of the hysteresis cycles.

Figure 4b shows the force/time curves obtained from the relaxation tests. We can observe the decrease of force against time for each sample at all times in its maturation process. Thus, the results show that the viscosity effect.

**Tensile experiments under quasi static loads:**

The mechanical properties of regenerated tissue were deduced from the force/displacement curves. These showed three phases (Figures 5).

The first non-linear part of the curve corresponds to the toes region. On the second part of the curve the behaviour is linearly elastic. On the third part, one can observe an increasing damage part (weakly non-linear) after which failure occurs suddenly.

Two modes of failure were observed (Figure 5):

- At 7 days, the regenerated tissue was soft and failure occurred at the interface of the bone and regenerated tissue. The tissue was damaged by delamination.
- From Day 14 and 21, the regenerated tissue ruptured sharply like a rigid material.

The evolution of the mean failure force and displacement value in all the samples is given in Table 1. The ultimate force $f_{rup}$ and displacement, $d_{rup}$ varied non-linearly between 7 and 21 days (Cf. Table 1)..

3 -3) Statistics:

The results show that it exists a significant correlation between failure force at day 7 and day 14 (Z=-2.65, p=0.01), at day 7 and day 21 (Z=-2.23, p=0.02) and at day 14 and day 21 (Z=-2.12, p= 0.03). Concerning failure displacement, we have a significant correlation only between day 7 and day 14 (Z=-2.65, p=0.01).

3 - 4) Qualitative histology:

The histological findings revealed enchondral ossification. Mesenchymal progenitors filled the tissue gap before Day 7 (Figure 6).

- At Day 7, the regenerating tissue was formed by heterogeneous tissues with a cartilaginous predominance. The central part was composed of 80% cartilage and 20% calcified cartilage turning into bone by enchondral ossification. The edges of the regeneration, in continuity with native bone, appeared more mature, with about 50% mineralization of the extracellular matrix.

- At Day 14, the central part of the regenerating tissue was composed of 50% cartilage and 50% of mineralised structures. Ossification seemed to occur along a mineralization gradient from the healthy bone edges towards the middle third of the newly produced tissue.

- At Day 21, in the central part of the regenerate 95% of the cartilage was mineralized.

## 4) Discussion

This experimental model enabled the regeneration of all the diaphyseal bone segment from the $7^{th}$ postoperative day. In order to obtain samples of regenerating tissue suitably long for mechanical testing, we removed 25 mm of the tibial diaphysis. The periosteum was left *in situ* in order to contribute quantitatively and qualitatively to the regeneration process. Indeed, its release from the bone surface triggers the proliferation of the mesenchymal progenitors from its cambium layer([2;14]) as well as the emission of growth factors which recruit precursor cells from the neighbour tissue and orientate their differentiation ([16-18]). Thus, the regenerative process lead to the production of different skeletal elements: bone cartilage or fibrous tissue, according to the environmental variables ([3]).

In this study, we sought to regenerate bone with minimal distortion of the spontaneous repair process. In particular we decided not to use any pharmacological intervention but to control the mechanical environment, as bone differentiation is promoted by low hydrostatic pressures and low strains ([4]). Tissue strains over 15% do not lead to bone formation, but rather favour fibrous tissue differentiation ([5]). In order to control the mechanical environment of the regenerating tissue during the surgery, we stabilized the tibiae externally. Due to the relative rigidity of the device and its anchorage with half pins, we assume that the regenerate was not subject to axial strain above this threshold. Whatever type of mono-lateral fixation device, some intrinsic flexibility remains, and some low, cyclic, axial compression and bend loadings may have occurred during locomotion. However, we consider these loadings were negligible. Thus, the basic level of tissue maturation produced gradual changes in tissue structure and mechanical behaviour according to genetically programmed time sequence. This study aimed to characterize their temporal evolution. In the literature, such studies are restricted to fracture and osteotomy models, which represent most typical clinical situations. The structural properties of the callus were tested by bending, compression, torsion or traction, at different time points of the healing process ($^{5;19;20,\ 21}$). These studies have been enhanced by numerical modelling ($^{5;12;13;21-24}$). However, the proximity of the two bone edges, separated by a short length of callus, where different tissues types coexist, is responsible for the histological heterogeneity. Thus, the global inter-fragmentary mechanical loading creates heterogeneous fields of strains which render these analyses very complex ($^{5;24}$).

Our approach differed clearly from the fracture model by the larger volume of the regeneration. The width of the segmental bone removed has two major consequences:

In a fracture model, the residual micro-motion that might persist between bone extremities 2 to 3 mm apart may represent significant strains. By increasing ten fold the length of the inter-fragmentary gap (25 mm), the amount of tissue strain was reduced by the same factor. This minimal tissue strain persuaded us to regard the regenerating tissues as unloaded.

Furthermore, the large inter-fragmentary gap surrounded by the periosteal sleeve created the conditions of a biological regeneration chamber, offering similar biological environment throughout most of the regenerating tissue.

The originality of our approach has been to obtain, over a period, measurements of the regenerating tissues' material properties, up to primary bone formation. Cyclic, relaxation and failure tests were carried out on the same sample because the cycling and relaxation tests were done on the elastic part of the response curve. We can conclude that viscosity and plasticity cannot be disregarded when describing the behaviour of regenerated tissue.

In order to characterize the stiffness of the regenerating structure, we carried out traction tests. We chose these mechanical loading conditions to avoid problems of buckling in compression of the soft tissues at Day 7. In the literature, most authors studying cartilage carried out compression tests which provided different mechanical parameters ([25-27]). Very few authors have realized traction tests ([28;29]) and, the tissue we obtained at Day 7 was already more rigid indicating that it was more mineralized than cartilage. The histological analysis demonstrated that the new production of tissue was then 80% cartilage, but enchondral ossification had already started, with 20% calcified cartilage and woven bone formed.

At Day 14, as the enchondral ossification of the regenerating cartilage anlages went forwards, we observed a much more rigid behaviour of the samples. At that time, the mineralization of the cartilage matrix and its ossification had reached 50%.

At Day 21 of regenerated tissue and healthy bone had a similar rigidity (figure 5), with similar representative curve. Nevertheless, the ossified regenerating structure appeared as woven bone, since the remodelling process had not yet occurred, whereas the healthy bone was constituted of cortical bone from the mid-diaphysis of the tibia.

This study had some limitations:

First, we have analyzed only a small number of samples at each time point. However, the behaviour of samples at 7, 14 and 21 days were qualitatively comparable.

Second, with the earliest regenerating structures in our study, failure at traction appeared at the interface between the regenerating tissue and bone by gradual delamination. Thus this study characterizes the structural response of the regenerating tissues rather than material properties of the whole bone.

Third, mechanical tests were performed on formaldehyde-fixated bones. The formaldehyde affects the mechanical properties and rigidify the structure ([15]). But, this article is primarily a method paper. Moreover, all the samples were put in the same conditions so we had the same effect in all the samples and we could compare them qualitatively without problem.

Our challenge was, secondarily, to define an overall behavioural feature to put into a numerical model. To complete the study, and in particular to avoid problems of delamination at the junction of soft and hard tissues, we will add other material characterization experiments such as indentation to these experiments. These experiments will provide information on the evolution of the material properties, particularly useful in the early stages of the regenerate maturation.

Conclusion:

This *in vivo* study demonstrated that it is possible to regenerate an entire bone volume after a wide segmental defect in the New Zealand rabbit. It analyzed the time sequences of the regeneration process without the influence of mechanical loading.

This experimental model initiates a cascade of cellular and molecular events whose determinism, in the biological and mechanical environment encountered, is bone tissue regeneration. These biological events are reproduced in a number of physiological and pathological conditions involving bone repair, such as unloaded fracture healing or implant integration. Thus, this study model is only an example and its results can be extrapolated to the different conditions which involve unloaded bone regeneration.

The current experiment can be considered as a control study which will allow comparison with the evolution of the biomechanical properties of regenerating skeletal tissue obtained in other environments. For instance, the influence of mechanical loading history, or the interaction between skeletal biomaterials and recipient bone could be compared with the evolution of the spontaneous process of skeletal tissue regeneration.

## Acknowledgements:

We would like to thank Unimeca (Marseilles) and particularly C. Hochard who allowed us to use the traction device.

We would like to thank M. Tellache for his contribution to the mechanical testing.

We would like to thank Dr. Pourquier for histological analyses.

We would like to thank Dr. J.L. Ferran and the staff of the radiology department at the Clinique St Jean (Montpellier) for their contribution to the imaging studies.

We would like to thank P. Chavet for her help in the statistical study.

We also thank S. Séguinel for her grammar and vocabulary corrections.

**Table legends**

Table 1: failure force ($f_{rup}$), and failure displacement ($d_{rup}$) variations obtained using quasi static devices

|  | 7 days | 14 days | 21 days | Healthy bone |
|---|---|---|---|---|
| **$f_{rup}$(N)** | $20 \leq f_{rup} \leq 50$ | $60 \leq f_{rup} \leq 100$ | $250 \leq f_{rup} \leq 350$ | 341 |
| **$d_{rup}$ (mm)** | $2.4 \leq d_{rup} \leq 8.3$ | $7.3 \leq d_{rup} \leq 9.6$ | $7{,}3 \leq d_{rup} \leq 11$ | 7.5 |

Table 1

# Figure legends

Figure 1: Representation of the 25 mm segmental bone removal in the mid-diaphysis of rabbit tibia, with osteosynthesis by external fixator and closure of the preserved periosteum.

Figure 2: Sample in the traction device with the extensometer.

Figure 3: Scanner images of explanted bone samples at Day 7 (a), Day 14 (b) and Day 21 (c). Angular malalignement can be measured accurately in the plane of deformity. Gradual mineralization of the regenerating bone is measured by densification of the regenerated area.

Figure 4: (a) is one example of a hysteresis curve during three cycles at day 21 on one sample. These curves can be divided in three parts. The first one corresponds to the loading. The second part is the relaxation and the third part represents the unloading. Hysteresis is clockwise. (b) is one example of relaxation curve at day 21.

The hysteresis and relaxation curves are qualitatively comparable at day 7, 14 and 21.

Figure 5: Quasi static force/displacement curves for all the samples at 7 days (in black) (a), 14 days (in black) (b), 21 days (dotted line) and for the healthy bone (continuous line) (c).

Figure 6: Histology of the regenerating tissues. Magnified x 40. The volume of the removed bone is restored by a cartilaginous anlage undergoing enchondral osteogenesis (a) Day 7, 20% of the area is composed of calcified cartilage (red trabeculae). (b) Day 14, 50% of the regenerated area is mineralised. (c) Day 21, 95% of the structure is mineralized. The calcified cartilage is gradually replaced by woven bone.

**Figures**

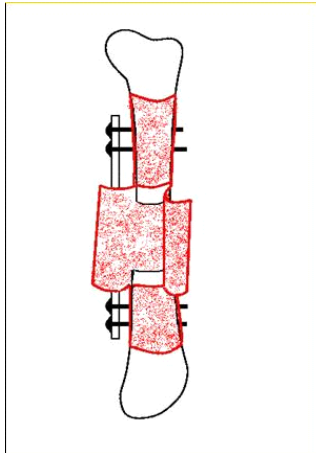
Figure 1

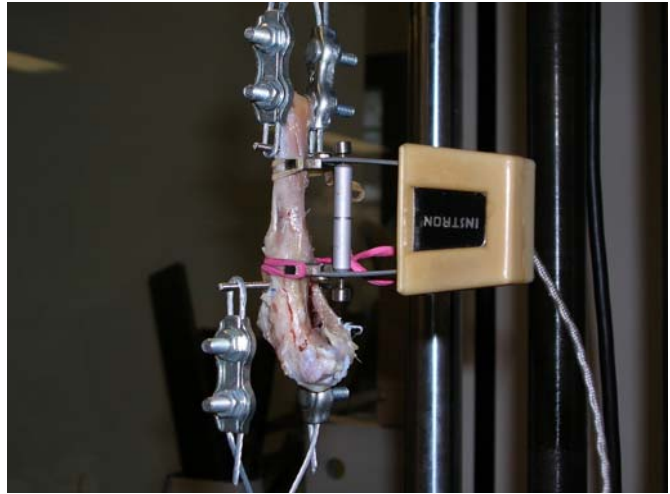

Figure 2

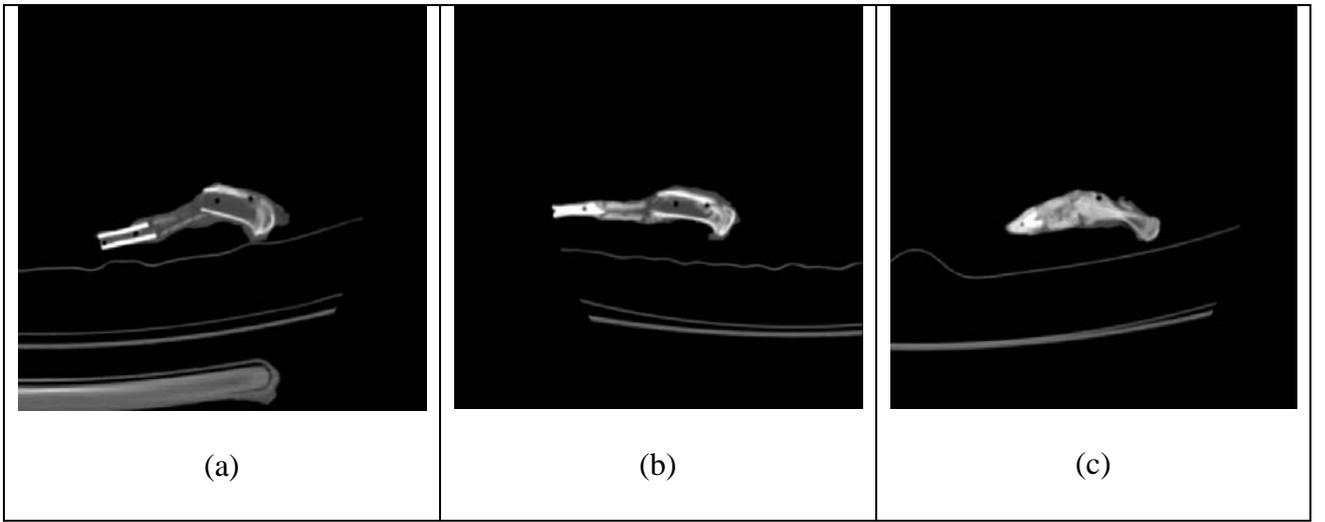

(a) (b) (c)

Figure 3

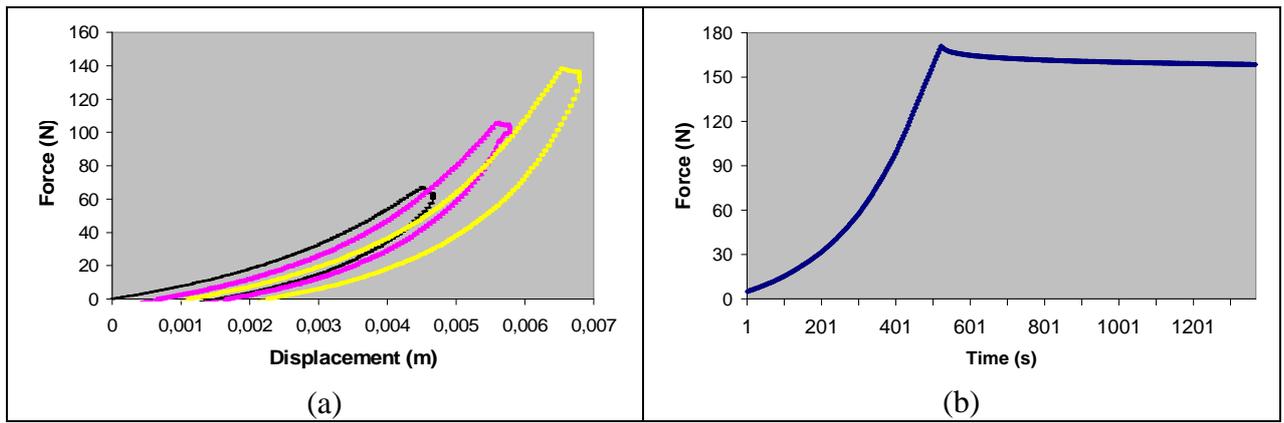

Figure 4

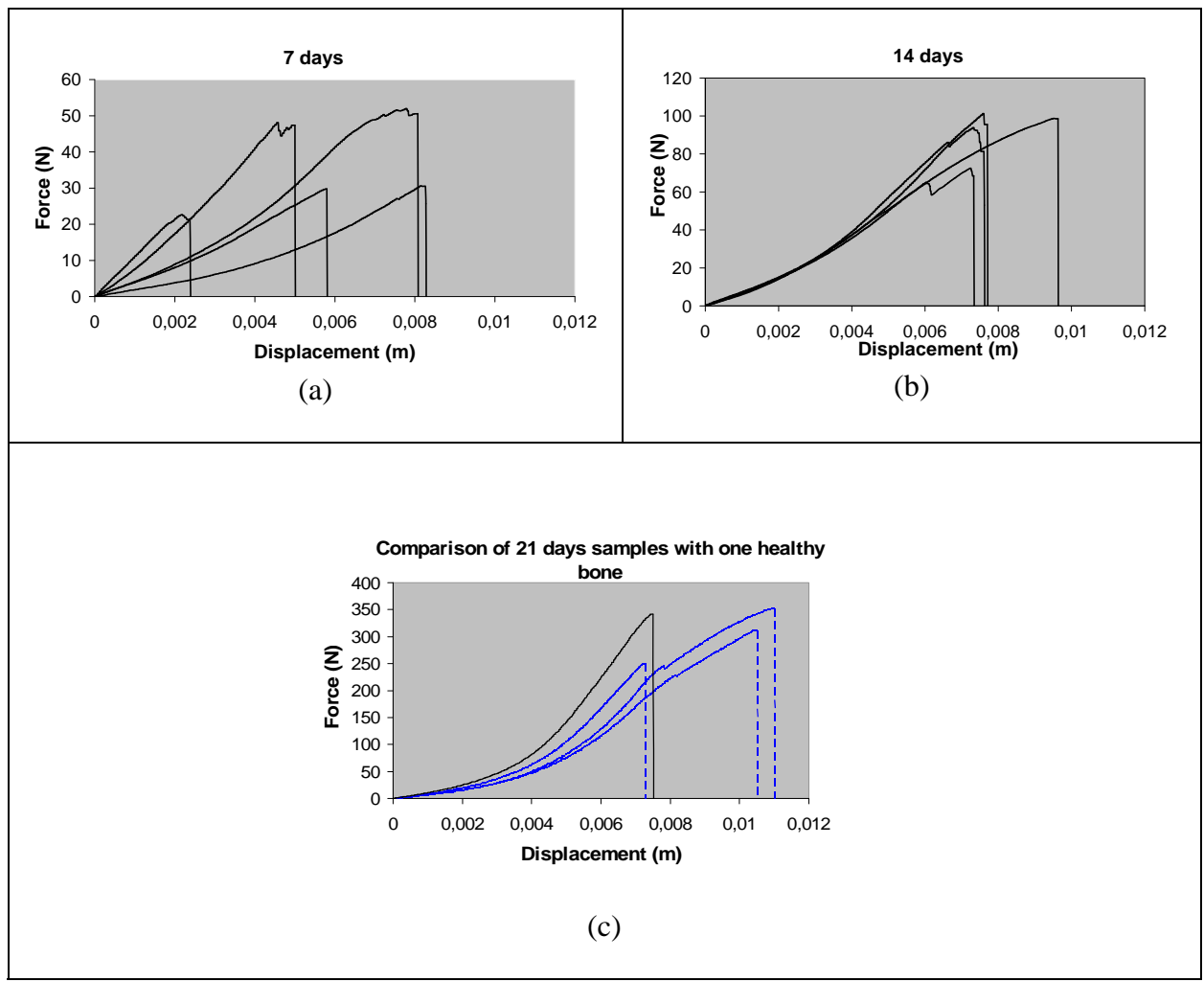

Figure 5

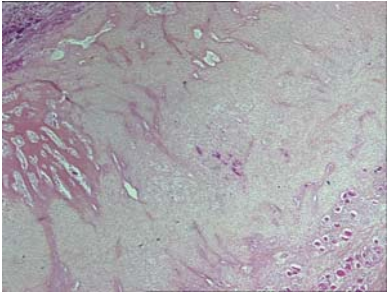 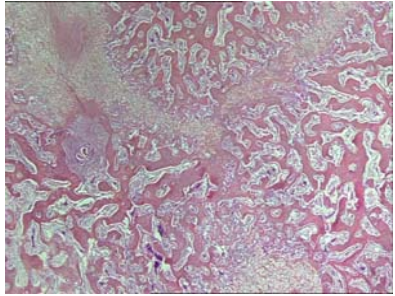 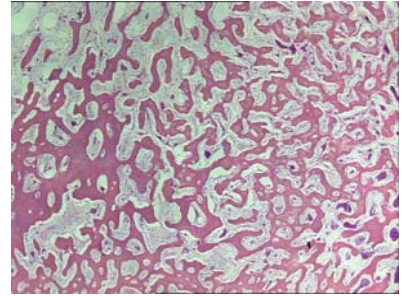

(a)                  (b)                 (c)

Figure 6